\documentclass{PoS}  
   \usepackage{amssymb}
   \usepackage{amsmath}
    \usepackage{amsfonts}
     \usepackage{epsfig}
      \usepackage{bm}

\usepackage{mathpazo}

\title{Dipole model analysis of the new HERA I+II data}

\ShortTitle{Dipole model analysis of the new HERA data}

\author{\speaker{Agnieszka Luszczak}%
        \thanks{This work is supported by the Polish Ministry under program Mobility Plus, no. 1320/MOB/IV/2015/0}\\
       Institute of Physics, T.Kosciuszko Cracow University of Technology, Poland\\
       E-mail: \email{agnieszka.luszczak@desy.de}}

\abstract{We use the dipole model to analyze the inclusive DIS  cross section data, obtained from the HERA I+II measurements \cite{Abramowicz:2015mha}. 
We show that these combined data are very well described within the dipole model framework, which is  complemented with a valence quark structure functions. 
Our motivation is  to investigate the gluon density with the BGK dipole model \cite{BGK} as an alternative to 
the PDF approach. BGK dipole model uses for evolution the DGLAP mechanism in the $kt$ factorization scheme (in contrast 
to the collinear factorization for PDFs). We confirm our results from the previous paper \cite{Luszczak:2013rxa} with old HERA data \cite{HERA2010}.
In addition we also performed a first, preliminary investigation of  saturation. The analysis was done in  the xFitter framework \cite{xFitter,xFitter2,xFitter3,xFitter4,xFitter5}.}

\FullConference{XXIV International Workshop on Deep-Inelastic Scattering and Related Subjects\\
		 11-15 April, 2016\\
		 DESY Hamburg, Germany}

\begin{document}

\section{Introduction}
HERA inclusive and diffractive DIS cross sections are very well described by the dipole models~\cite{BGK,GBW, Iancu:2003ge,Forshaw:2006np}. 
Interest in the dipole description emerge from the fact that dipole picture provides a natural description
of QCD reaction in the low $x$ region. Due to the optical theorem, 
dipole models allow a simultaneous description of many different physics reactions, 
like inclusive DIS processes, inclusive diffractive processes, exclusive $J/\psi$, $\rho, \phi$ production, diffractive jet production, 
or diffractive and non-diffractive charm production. In the dipole picture, all these processes are determined by the same, universal, 
gluon density~\cite{Shoshi:2002in,MSM,KT,KMW,Watt:2007nr}.  
  The direct connection between the dipole production and gluon density  is particularly clearly seen in the exclusive $J/\psi$ production. 
 Another important application of the dipole description is the investigation of the gluonic high density states. These can be characterized
  by the degree by which a dipole is absorbed or multiply scattered in such states. The states with the highest gluon densities are produced today in the high energy  heavy ion scattering at RHIC and LHC. 
\section{Formalism}

In the dipole picture the deep inelastic scattering is viewed as a two stage process; 
first the virtual photon fluctuates into a dipole,
which consists of a quark-antiquark pair (or a $q\bar q g$ or $q\bar q gg$ ... system) 
and in the second stage the dipole interacts with the proton.
  The scattering amplitude is a product of the virtual photon wave function, $\Psi$, 
  with the dipole cross section, $\sigma_{\text{dip}}$, which determines a probability of the dipole-proton scattering. 
Thus, within the dipole formulation of the $\gamma^* p$ scattering we write:  
\begin{equation}
   \sigma_{T,L}^{\gamma^* p}(x,Q^{2}) = \int d^2r \int dz \Psi^*_{T,L} (Q,r,z) \sigma_{\text{dip}}(x,r) \Psi_{T,L}(Q,r,z),
\end{equation}
where $T,L$ denotes the virtual photon polarization, $\sigma_{T,L}^{\gamma^* p}$ the total inclusive DIS cross section.
In the analysis presented here we concentrate on BGK dipole model. In this model the evolution ansatz of the GBW model was improved in the model proposed by Bartels, Golec-Biernat and Kowalski, (BGK)~\cite{BGK},  by taking into account the  DGLAP evolution of the gluon density in an explicit way. The model preserves the GBW eikonal approximation to saturation and thus the dipole cross section is given by:
\begin{equation}
\label{eBGK}
   \sigma_{\text{dip}}(x,r^{2}) = \sigma_{0} \left(1 - \exp \left[-\frac{\pi^{2} r^{2} \alpha_{s}(\mu^{2}) xg(x,\mu^{2})}{3 \sigma_{0}} \right]\right).
\end{equation}
The evolution scale $\mu^{2}$ is connected to the size of the dipole by $\mu^{2} = C/r^{2}+\mu^{2}_{0}$. 
The gluon density, which  is parametrized  at the starting scale $\mu_{0}^{2}$, 
is evolved to larger scales, $\mu^2$, using LO or NLO DGLAP evolution.
 We consider here gluon density in the following form:
\begin{equation}
   xg(x,\mu^{2}_{0}) = A_{g} x^{-\lambda_{g}}(1-x)^{C_{g}},
\end{equation} 
The free parameters for this model are $\sigma_{0}$, $\mu^{2}_{0}$ and the three parameters for gluon $A_{g}$, $\lambda_{g}$, $C_{g}$. 
 Their values are obtained by a fit to the data. The fit results were found to be independent on the parameter $C$, which was therefore fixed as  $C=4$ GeV$^2$, in agreement with the original BGK fits.


\section{Results from fits}
In the Table~\ref{tabl1} we show dipole model BGK fits with valence quarks. 
\begin{table}[ht]
\begin{center}
\begin{tabular}{|c|c|c|c|c|c|c|c|c|c|c|c|} 
\hline 
No& 
$Q^2$&
$\sigma_0$ & $A_g$ & $\lambda_g$ & $Cg$ &$cBGK$& $Np$& $\chi^2$& $\chi^2/Np$\\
\hline
1 &
$Q^2 \ge 3.5 $ & 85.111 &  1.857&  -0.12596&  11.339 &4.0& 568 & 605.29 &  1.07 \\
\hline
2 &
$Q^2 \ge 8.5$ & 72.451 & 2.015 & -0.1185 & 12.682 &4.0&  482 &  495.44 &  1.03 \\
\hline
\end{tabular}
\end{center}
\caption{BGK NLO fit with valence quarks for $\sigma_r$
for HERA1+2-NCep-460, HERA1+2-NCep-575,   HERA1+2-NCep-820, 
HERA1+2-NCep-920 and HERA1+2-NCem in the range $Q^2 \ge 3.5$ GeV$^2$ and $Q^2 \ge 8.5$ and $x\le 0.01$. {\it Soft gluon}.}
\label{tabl1}
\end{table}

In the Table~\ref{tabl2} we show dipole model BGK fits without valence quarks.
\begin{table}[hb]
\begin{center}
\begin{tabular}{|c|c|c|c|c|c|c|c|c|c|c|c|} 
\hline 
No& 
$Q^2$&
$\sigma_0$ & $A_g$ & $\lambda_g$ & $Cg$ &$cBGK$& $Np$& $\chi^2$& $\chi^2/Np$\\
\hline
1 &
$Q^2 \ge 3.5 $ & 85.111 &  2.075&  -0.093&  4.989 &4.0& 568 & 592.46 &  1.04 \\
\hline
2 &
$Q^2 \ge 8.5$ & 123.31 & 1.997 & -0.0975 & 4.655 &4.0&  482 &  479.37 &  0.99 \\
\hline
\end{tabular}
\end{center}
\caption{BGK NLO fit without valence quarks for $\sigma_r$  for HERA1+2-NCep-460, HERA1+2-NCep-575,   HERA1+2-NCep-820, HERA1+2-NCep-920 and HERA1+2-NCem in the range $Q^2 \ge 3.5$ GeV$^2$ and $Q^2 \ge 8.5$ and $x\le 0.01$. {\it Soft gluon}.}
\label{tabl2}
\end{table}

In the Table~\ref{tabl3} we show dipole model BGK fits with fitted valence quarks. 
\begin{table}[hb]
\begin{center}
\begin{tabular}{|c|c|c|c|c|c|c|c|c|c|c|c|} 
\hline 
No& 
$Q^2$&
$\sigma_0$ & $A_g$ & $\lambda_g$ & $Cg$ &$cBGK$& $Np$& $\chi^2$& $\chi^2/Np$\\
\hline
1 &
$Q^2 \ge 3.5 $ & 85.111 &  1.921&  -0.103&  4.674 &4.0& 557 & 575.30 &  1.03 \\
\hline
2 &
$Q^2 \ge 8.5$ & 93.581 & 1.665 & -0.124 & 6.066 &4.0&  473 &  476.71 &  1.01 \\
\hline
\end{tabular}
\end{center}
\caption{BGK NLO fit with fitted valence 
quarks for $\sigma_r$  for HERA1+2-NCep-460, HERA1+2-NCep-575,  
HERA1+2-NCep-820, HERA1+2-NCep-920 and HERA1+2-NCem in the range $Q^2 \ge 3.5$ GeV$^2$ and $Q^2 \ge 8.5$ and $x\le 0.01$. {\it Soft gluon}.} 
\label{tabl3}
\end{table}

In the Table~\ref{tabl4} we show HERAPDF standard fits. 
\begin{table}[hb]
\begin{center}
\begin{tabular}{|c|c|c|c|c|c|c|c|c|c|c|c|} 
\hline 
No& 
$Q^2$&HF Scheme&
$Np$& $\chi^2$& $\chi^2/Np$\\
\hline
1 &
$Q^2 \ge 3.5 $ & RT& 1131 &  1356.70 & 1.20 \\
\hline
2 &
$Q^2 \ge 8.5$ & RT&  456 &  470.88 & 1.15 \\
\hline
\end{tabular}
\end{center}
\caption{HERAPDF NLO fit with fitted valence quarks for $\sigma_r$ for HERA1+2-NCep-460, HERA1+2-NCep-575  HERA1+2-NCep-820, HERA1+2-NCep-920, HERA1+2-NCem, HERA1+2-CCep and HERA1+2-CCem data in the range $Q^2 \ge 3.5$ and $Q^2 \ge 8.5$ and $x\le 1.0$.}
\label{tabl4}
\end{table}


In the Table~\ref{tabl5} we show dipole model BGK fits without saturation, when dipole cross section is in the form: 
\begin{equation}
\sigma(r,x) = \sigma_0 ~ [\pi^2r^{2}\alpha_s(\mu^2)xg(x,\mu^2)/(3\sigma_0)]
 \end{equation}
 
\begin{table}[ht]
\begin{center}
\begin{tabular}{|c|c|c|c|c|c|c|c|c|c|c|c|} 
\hline 
No& 
$Q^2$&
$\sigma_0$ & $A_g$ & $\lambda_g$ & $Cg$ &$cBGK$& $Np$& $\chi^2$& $\chi^2/Np$\\
\hline
1 &
$Q^2 \ge 3.5 $ & 118.12 &  1.0535&  -0.103&  -0.389 &4.0& 568 &874.72 & 1.54 \\
\hline
2 &
$Q^2 \ge 8.5$ & 118.12 & 0.914 & -0.130 & -0.484 &4.0&  482 &592.86 & 1.23 \\
\hline
\end{tabular}
\end{center}
\caption{BGK NLO fit without valence quarks for $\sigma_r$  for HERA1+2-NCep-460,
HERA1+2-NCep-575,   HERA1+2-NCep-820, HERA1+2-NCep-920 and HERA1+2-NCem in the range $Q^2 \ge 3.5$ GeV$^2$ and
$Q^2 \ge 8.5$ and $x\le 0.01$. {\it Soft gluon}.}
\label{tabl5}
\end{table}
When we compare BGK fits from Table~\ref{tabl5} without saturation with proper fits from Table~\ref{tabl2} but with saturation: 
\begin{equation}
 \sigma(r,x) = \sigma_0 \left\{1-\exp\left[-\pi^2r^{2}\alpha_s(\mu^2)xg(x,\mu^2)/(3\sigma_0)\right]\right\}
\end{equation} 
we see that fits from Table~\ref{tabl5} give much worse $\chi^2/Np$.

\section{Comparision with HERA I+II data}

We obtained very good description of HERA I+II data using BGK dipole model for every $Q^2$. Here we show the comparision of our fit with data for a few selected $Q^2$.
\begin{figure*}[hb]
 \begin{center}
\includegraphics[width=10.0cm]{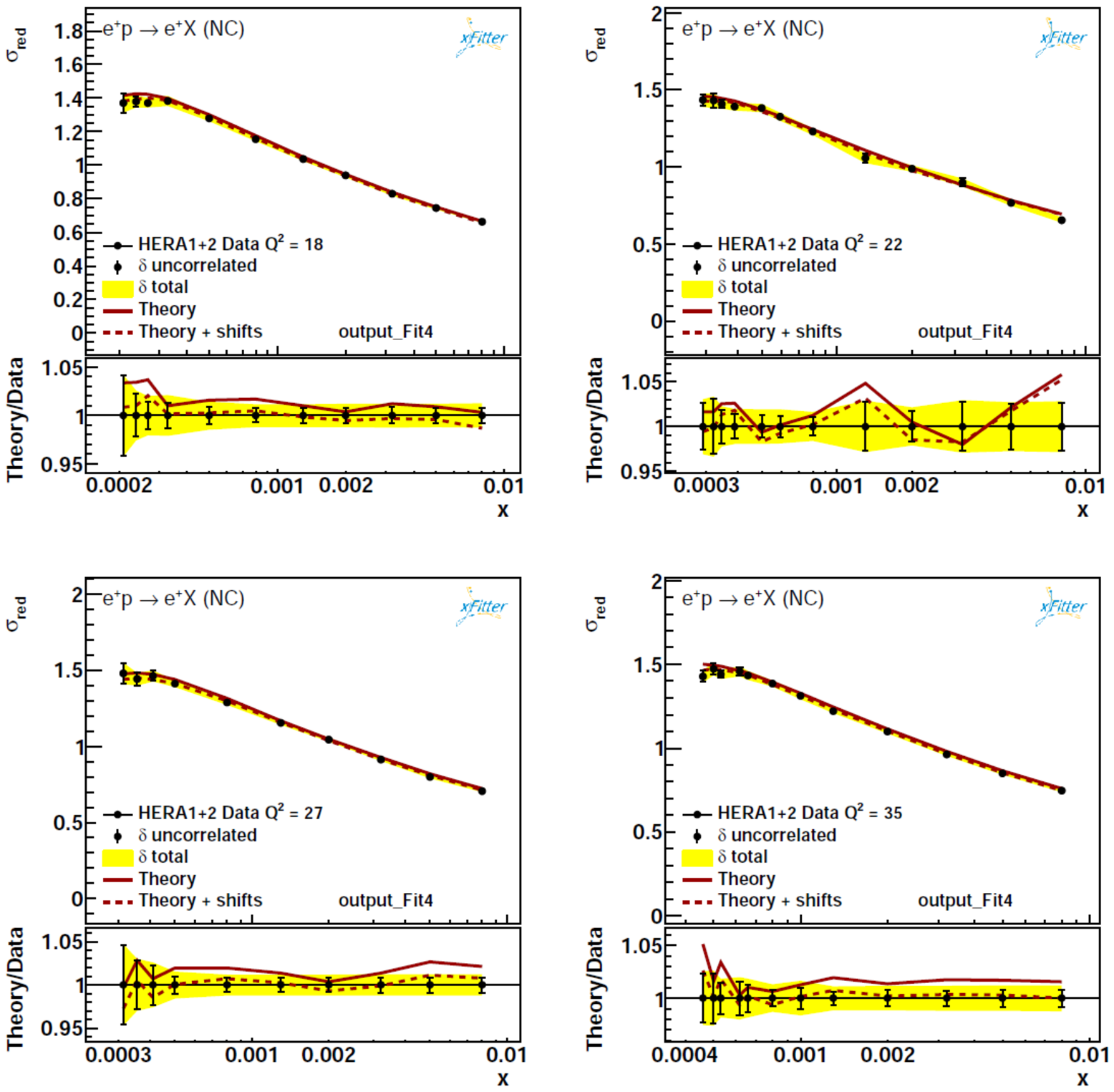}
\end{center}
\caption{Comparision of BGK dipole fit Nb.1 from Table 3 with HERA data.}
\end{figure*}

\section{Conclusions}

We have shown that BGK dipole fits (with saturation) describe the final, high precision HERA I+II  data 
with $x< 0.01$, very well. We observed the little sensitivity to valence quarks contribution. 
BGK dipole model fits without saturation give a much worse $\chi^2/Np$. We examine this issue.

\section{Acknowledgement}
I would like to thank H.Kowalski (DESY) and S.Glazov (DESY) for useful comments to the analysis presented here.

 
\end{document}